\documentclass
[tightenlines,final,letterpaper,10pt,pre,amsfonts,amssymb,twocolumn,showkeys,byrevtex,preprint,noshowpacs]{revtex4}%
\usepackage{amsfonts}
\usepackage{amsmath}
\usepackage{amssymb}
\usepackage{graphicx}%
\providecommand{\U}[1]{\protect\rule{.1in}{.1in}}

\begin{document}
\title{Loading-unloading hysteresis loop of randomly rough adhesive contacts.}
\author{Giuseppe Carbone}
\affiliation{Department of Mechanics, Mathematics, Management, Politecnico di Bari, v.le
Japigia 182, I-70126, Bari - Italy}
\affiliation{CNR - Institute for Photonics and Nanotechnologies U.O.S. Bari, Physics
Department M. Merlin, via Amendola 173, I-70126 Bari, Italy}
\author{Elena Pierro}
\affiliation{Scuola di Ingegneria - Universit\`{a} degli Studi della Basilicata, Campus di
Macchia Romana - Via dell'Ateneo Lucano 10, I-85100 Potenza - Italy}
\author{Giuseppina Recchia}
\affiliation{Department of Mechanics, Mathematics, Management, Politecnico di Bari, v.le
Japigia 182, I-70126, Bari - Italy}
\begin{abstract}
In this paper we investigate the loading and unloading behavior of soft solids
in adhesive contact with randomly rough profiles. The roughness is assumed to
be described by a self-affine fractal on a limited range of wave-vectors. A
spectral method is exploited to generate such randomly rough surfaces. The
results are statistically averaged, and the calculated contact area and
applied load are shown as a function of the penetration, for loading and
unloading conditions. We found that the combination of adhesion forces and
roughness leads to a hysteresis loading-unloading loop. This shows that energy
can be lost simply as a consequence of roughness and van der Waals forces, as
in this case a large number of local energy minima exist and the system may be
trapped in metastable states. We numerically quantify the hysteretic loss and
assess the influence of the surface statistical properties and the energy of
adhesion on the hysteresis process.

\end{abstract}
\maketitle

\section{Introduction}

Contact mechanics between rough surfaces plays a crucial role in a large
number of engineering applications, ranging from seals \cite{Persson-2008,
Lorenz-2010}, boundary and mixed lubrication \cite{Scaraggi-2011,
Scaraggi-2011BIS}, adhesive systems and friction \cite{Carbone-2008,
Zhao-2003, Carbone-2013}. Recently, an increasing interest in these topics has
been motivated by the increasing efforts made to face up new technological
challenges, such as the manufacturing of novel bio-inspired adhesives
\cite{Geim-2003, Carbone-2011, Carbone-2012}, the optimization of
seals\ \cite{Bottiglione-2009, Dapp-2012}, the extreme downsizing of
mechanical and electrical devices. In particular, micro- and nano-mechanical
systems (MEMS/NEMS) have driven the development of new materials and surfaces,
involving an increasing influence of surface phenomena. In such applications,
where usually surfaces may experience rapid intermittent contacts, it is then
very important to know how adhesion and roughness affect the behavior of the
system and in particular the energy dissipation at the interface. This aspect
of the problem is also very critical when scanning probe microscopy, as atomic
force microscopy (AFM), is utilized to characterize the mechanical behavior
and surface properties of materials. During adhesive contacts of rough solids
the measured contact force versus displacement shows a clear hysteretic loop
associated with energy dissipation \cite{Pickering-2001, Chen-1991,
Maeda-2002}, which is not observed in perfectly smooth elastic contacts.

The relevance of the problem have strongly stimulated this research field.
However, notwithstanding the quite large amount of papers dealing with the
adhesive contact of rough surfaces \cite{FullerTabor-1975, Pastewka-2014,
Hyun-2004, Luan-2009, Jackson-2011, Medina-2014, Medina-2013, Sellgren-2003},
only a few papers attempt to explain the origin of adhesion hysteresis and
energy dissipation in rough contacts \cite{Li-2015, Kadin-2006, Kadin-2008}.
This is usually attributed to the presence of plasticity \cite{Eid-2011,
Kadin-2006, Kadin-2008}, interdigitations among polymer chains
\cite{Maeda-2002}, humidity \cite{Pickering-2001}, viscoelasticity
\cite{Tirrell-1996, Greenwood-1981}. In fact all these phenomena can act
contemporaneously, so that distinguishing among the different causes has
become of utmost importance. Only a few works have made an effort in such
direction. For example Refs. \cite{Guduru-2007, Guduru-2007BIS, Zappone-2007,
Kesari-2010} proved that adhesion hysteresis can be observed even in the case
of elastic solids provided that the contact occurs between rough surfaces.The
study presented herewith aims to provide an additional contribution along this
direction. By employing a methodology based on a pure continuum mechanics
approach, which belongs to the class of Boundary Element Methods (BEMs)
\cite{Carbone-2008, Carbone-2012, Carbone-2009}, we analyse numerically the
loading-unloading adhesive contact of rough solids by including adhesion in
terms of surface energy, i.e. by assuming that the range of the adhesive
interaction is infinitely short. We study, in particular, the influence of
adhesion and surface roughness on the hysteresis loop, and show that energy
can be lost simply as a consequence of roughness and van der Waals forces.
Such energy loss is numerically quantified and explained in terms of two
different but concurrent mechanisms occurring at the contact interface and
involving different ranges of roughness length scales.

\section{The numerical model}

\begin{figure}[h]
\begin{center}
\includegraphics[width=8.00cm]{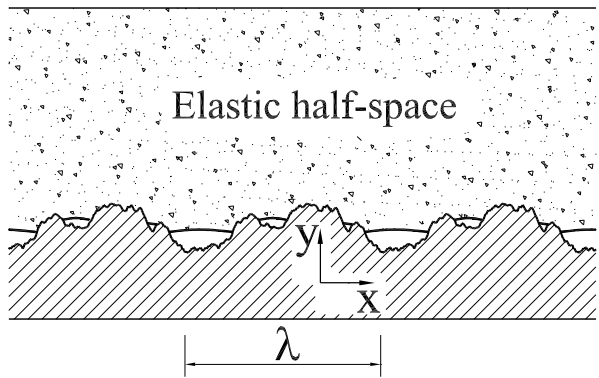}
\end{center}
\caption{An elastic half-space in contact with a periodic randomly rough rigid
substrate of wavelength $\lambda$.}%
\label{01}%
\end{figure}

\begin{figure}[h]
\begin{center}
\includegraphics[width=8.00cm]{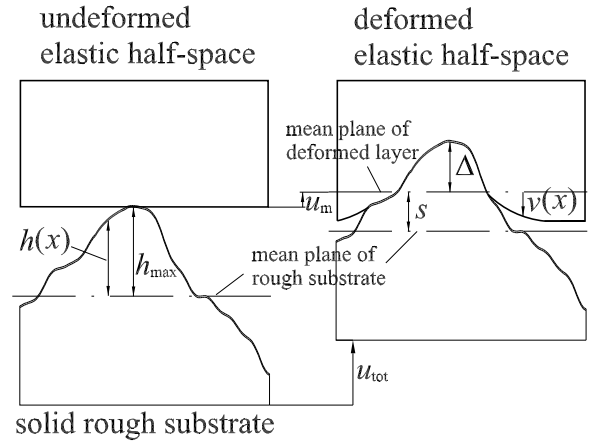}
\end{center}
\caption{A schematic representation of elastic displacement field at the
interface, as they occur during approaching the substrate to the elastic solid
by a quantity $u_{\mathrm{tot}}$. Observe that $u_{\mathrm{tot}}$ is the sum
of the mean displacement $u_{\mathrm{m}}$ of the elastic body and substrate
penetration $\Delta$. Also the interfacial displacement and the separation $s$
between the two surfaces are shown.}%
\label{02}%
\end{figure}

We briefly summarize the numerical methodology presented in
\cite{Carbone-2008, Carbone-2012, Carbone-2009}. We consider a periodic
problem where an elastic half-space is in contact with a randomly rough rigid
profile of height distribution $h\left(  x\right)  $ as shown in Fig.
\ref{01}, where $\lambda$ is the spatial period of the profile. The quantity
$h\left(  x\right)  $ is the height of the profile measured from its mean
plane. Because of periodicity $h\left(  x\right)  $ can be represented in
exponential form as%
\begin{equation}
h\left(  x\right)  =%
{\displaystyle\sum\limits_{m=-\infty}^{+\infty}}
a\left(  q_{m}\right)  e^{iq_{m}x}=%
{\displaystyle\sum\limits_{m=-\infty}^{+\infty}}
a_{m}e^{imq_{0}x} \label{general rough profile}%
\end{equation}
where the fundamental wavevector $q_{0}=2\pi/\lambda$, $m$ is the
wavenumber, $a_{m}=\left\vert a_{m}\right\vert e^{i\phi_{m}}$ and $\phi_{m}$
the phase of the $m$-th spectral component, uniformly distributed in the
interval $\left[  0,2\pi\right[  $. Figure \ref{02} shows the total
displacement $u_{\mathrm{tot}}$ of the substrate, the average displacement
$u_{\mathrm{m}}$ of the boundary of the deformed layer, and the penetration
$\Delta$ of the rigid substrate into the elastic layer. These three quantities
are shown to satisfy the following relation%
\begin{equation}
u_{\mathrm{tot}}=\Delta+u_{\mathrm{m}} \label{indenter displacement}%
\end{equation}
Figure \ref{02} also shows the so called interfacial displacement
$v\left(  x\right)  $, and the separation $s=h_{\max}-\Delta$ between the two
surfaces, where $h_{\max}=\max\left[  h\left(  x\right)  \right]  $ is the
maximum height of the roughness from its mean plane. Let us define the contact
domain $\mathrm{\Omega}=\cup_{i=1}^{L}\left[  a_{i},b_{i}\right]  $, where
$a_{i}$ and $b_{i}$ are the coordinates of $i$-th contact patch with
$a_{i}<b_{i}$ and $i=1,2,...,L$, where $L$ is the number of contacts patches.
Recalling that the interfacial pressure distribution $\sigma\left(  x\right)
$ vanishes out of the contact domain the relation between $\sigma\left(
x\right)  $ and the interfacial displacement $v\left(  x\right)  $ is%
\begin{equation}
-\int_{\mathrm{\Omega}}\mathcal{G}\left(  x-x^{\prime}\right)  \sigma\left(
x^{\prime}\right)  dx^{\prime}=v\left(  x\right)
\label{general green equatin}%
\end{equation}
where the kernel \cite{Carbone-2008, Carbone-2009}%
\begin{equation}
\mathcal{G}\left(  x\right)  =\frac{2\left(  1-\nu^{2}\right)  }{\pi E}%
\log\left[  2\left\vert \sin\left(  \frac{kx}{2}\right)  \right\vert \right]
\label{kernel function}%
\end{equation}
represents the Green function of the semi-infinite elastic body under a
periodic loading, i.e. it represents the interfacial displacement $v\left(
x\right)  $ caused by the application of a Dirac comb with peaks
$\delta\left(  x-n\lambda\right)  $ separated by a distance $\lambda$. Here
$E$ and $\nu$ are Young's modulus and Poisson's ratio of the elastic layer.
Now let us define the separation between the elastic solid and the rigid rough
substrate, i.e. the distance between the mean plane of the deformed surface
and the mean plane of the rough surface, as $s=h_{\max}-\Delta$ (see
Fig.\ref{02}). Noting that in the contact domain $v\left(  x\right)  =h\left(
x\right)  -s=h\left(  x\right)  -h_{\max}+\Delta$ one can also write%
\begin{align}
-\int_{\mathrm{\Omega}}\mathcal{G}\left(  x-s\right)  \sigma\left(  s\right)
ds  &  =h\left(  x\right)  -s;\qquad x\in\mathrm{\Omega}%
\label{integral equation}\\
-\int_{\mathrm{\Omega}}\mathcal{G}\left(  x-s\right)  \sigma\left(  s\right)
ds  &  =v\left(  x\right)  ;\qquad x\notin\mathrm{\Omega}
\label{integral equation bis}%
\end{align}
Eq. (\ref{integral equation}) is a Fredholm integral equation of the first
kind used to determine the unknown pressure distribution in the contact area
$\mathrm{\Omega}$. Whereas Eq. (\ref{integral equation bis}) is employed to
calculate the displacement $v\left(  x\right)  $ out of the contact area by
simply performing the integral at the left hand side. In order to close the
system of equations we need an additional condition to determine the yet
unknown contact domain $\mathrm{\Omega}$. To this end, we first observe that
for any penetration $\Delta$ or equivalently for any given separation $s$, we
can calculate the pressure distribution at the interface through Eq.
(\ref{integral equation}), and the interfacial elastic displacement through
Eq. (\ref{integral equation bis}), as functions of the unknown coordinates
$a_{i}$ and $b_{i}$ of the $i$-th contact area. To calculate the exact values
of the quantities $a_{i}$ and $b_{i}$ at equilibrium we need to minimize the
interfacial free energy $U_{\mathrm{tot}}\left(  a_{1},b_{1},...,a_{L}%
,b_{L},\Delta\right)  $ of the system at fixed penetration $\Delta$
\cite{Carbone-2008, Carbone-2009}.

The free interfacial energy is
\begin{equation}
U_{\mathrm{tot}}=U_{\mathrm{el}}+U_{\mathrm{ad}} \label{interfacial energy}%
\end{equation}
where the interfacial elastic energy $U_{\mathrm{el}}$ is \cite{Carbone-2008,
Carbone-2009}
\begin{equation}
U_{\mathrm{el}}\left(  a_{1},b_{1},...,a_{L},b_{L},\Delta\right)  =\frac{1}{2}%
{\displaystyle\sum\limits_{i=1}^{L}}
\int_{a_{i}}^{b_{i}}\sigma\left(  x\right)  \left[  h\left(  x\right)
-s\right]  dx. \label{interfacial elastic energy}%
\end{equation}
The adhesion energy is%
\begin{equation}
U_{\mathrm{ad}}\left(  a_{1},b_{1},...,a_{L},b_{L}\right)  =-\gamma%
{\displaystyle\sum\limits_{i=1}^{L}}
\int_{a_{i}}^{b_{i}}\sqrt{1+\left[  h^{\prime}\left(  x\right)  \right]  ^{2}%
}dx. \label{adhesion energy general}%
\end{equation}
where $\gamma$ is the work of adhesion. We, indeed, assume that the range of
adhesive interaction is infinitely short as in the JKR theory
\cite{Johnson-1971}. This assumption together with the law of
incompenetrability of bodies make the rigid wall behave as a bilateral
constraint, whose normal reaction forces per unit area may be either positive
(hard-wall repulsion) or negative (adhesive attraction). In this case the
stress field at the interface is only determined by enforcing the equilibrium
of the elastic body.

Eqs. (\ref{integral equation}), together with the requirement that the
interfacial free energy $U_{\mathrm{tot}}$ has a (local) minimum at
equilibrium, constitute a set of closed equations which allows, for any given
penetration $\Delta$, to determine the coordinates $a_{i}$ and $b_{i}$ of each
contact spot, the pressure distribution at the interface, and all other
physical quantities. For the numerical implementation the reader is referred
to \cite{Carbone-2009}.

The numerical simulations have been carried out for a randomly rough profile
with PSD%
\begin{align}
C_{R}\left(  q\right)   &  =C_{0}\left(  \frac{\left\vert q\right\vert
}{q_{\min}}\right)  ^{-\left(  2H+1\right)  };\qquad q\in\left[  q_{\min
},q_{\max}\right] \label{PSD}\\
C_{R}\left(  q\right)   &  =0;\qquad q\notin\left[  q_{\min},q_{\max}\right]
\nonumber
\end{align}
where $H$ is the Hurst exponent of the randomly rough profile. It is related
to the fractal dimension $D_{\mathrm{f}}=2-H$. In Eq. \ref{PSD} $q_{\min
}=n_{0}q_{0}$ and $q_{\max}=Nq_{\min}$. The generation of roughness has been
carried out by means of a spectral technique as shown in the Appendix.

\section{Results}

We assume that the elastic block is a soft perfectly elastic material with
elastic modulus $E=1\mathrm{MPa}$ and Poisson's ratio $\nu=0.5$. For each
rough profile results have been averaged over $10$ different realizations.

\begin{figure}[h]
\begin{center}
\includegraphics[width=8.00cm]{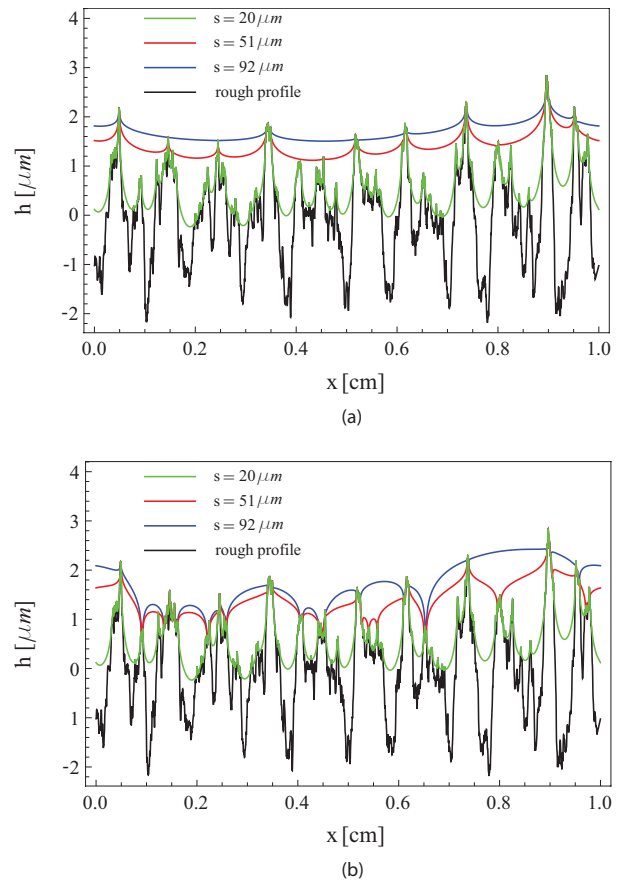}
\end{center}
\caption{(Color online) The deformed shapes of the elastic body at three
different separations, $s=92\mathrm{\mu m}$, $s=51\mathrm{\mu m}$, and
$s=20\mathrm{\mu m}$, and the rough rigid substrate profile, for a) loading
and b) unloading. The work of adhesion is $\gamma=0.01\mathrm{J/m}^{2}$, and
the rigid rough substrate profile has a fractal dimension $D_{\mathrm{f}}%
=1.2$.}%
\label{03}%
\end{figure}

\begin{figure}[h]
\begin{center}
\includegraphics[width=8.00cm]{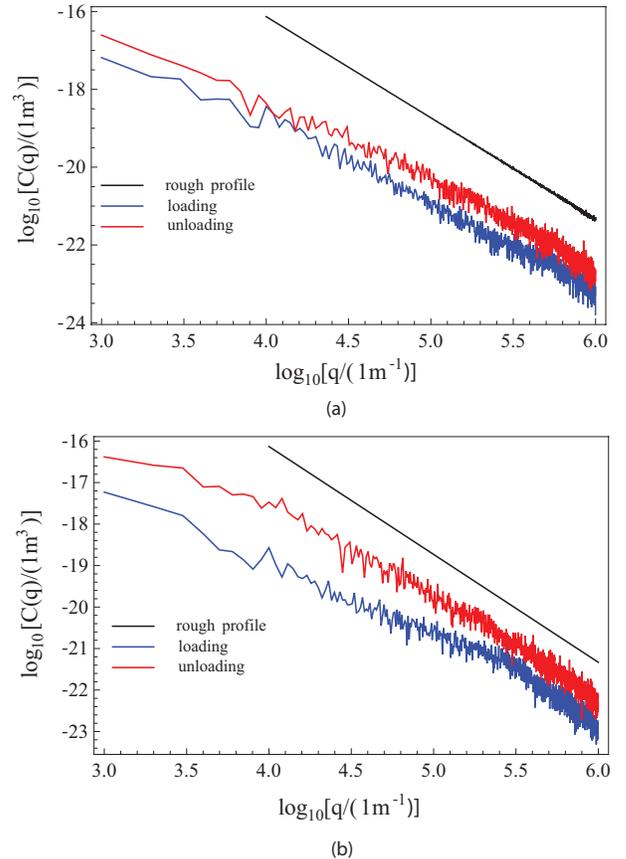}
\end{center}
\caption{(Color online) The PSD of the rigid substrate profile with fractal
dimension $D_{\mathrm{f}}=1.2$, compared to the PSD\ of the deformed shape of
the elastic body at fixed non dimensional penetration $\tilde{\Delta}%
=\Delta/h_{\max}=0.4$, obtained for (a) $\gamma=0.01\mathrm{J/m}^{2}$ and (b)
$\gamma=0.04\mathrm{J/m}^{2}$, for loading (blue curves) and unloading (red
curves) conditions.}%
\label{04}%
\end{figure}

The profiles have root mean square roughness $h_{\mathrm{rms}}=\left\langle
h^{2}\right\rangle ^{1/2}=1\mathrm{\mu m}$. The spectral components of our
profiles are given by Eq.\ref{PSD} and cover the wave-vector range from
$q_{\min}=n_{0}q_{0}$ up to $q_{\max}=Nq_{\min}$, outside this range the PSD
is zero. We have considered $\lambda=2\pi/q_{0}=6.28\mathrm{mm}$, $q_{\min
}=10~q_{0}$ and $q_{\max}=100~q_{\min}$. We note that once $h_{\mathrm{rms}}$,
$q_{\min}$ and $q_{\max}$ are fixed, changing the Hurst exponent (i.e. the
fractal dimension of the surface) also determines a modification of the
average square slope $m_{2}=$ $\left\langle h^{\prime2}\right\rangle =\int
q^{2}C_{R}\left(  q\right)  dq$ of the surface. For each generated rough
profile (see Sec. \ref{appendice}) the numerical calculations have been
carried out for different values of the separation $s=h_{\max}-\Delta$. In
Fig. \ref{03} we show three different shapes of the deformed profile at three
different values of the separation: $s=92\mathrm{\mu m}$, $s=51\mathrm{\mu m}%
$, and $s=20\mathrm{\mu m}$, for (a) loading and (b) unloading conditions. The
work of adhesion is $\gamma=0.01\mathrm{J/m}^{2}$. The rigid rough substrate
profile has a fractal dimension $D_{\mathrm{f}}=1.2$. In Fig. \ref{03} the
value $s=20\mathrm{\mu m}$ is the minimum value of separation at which the
unloading process begins to take place after loading. Therefore at
$s=20\mathrm{\mu m}$ the shape of the deformed profile is the same not
depending on what condition (i.e. loading or unloading) is being considered.
However, as the unloading proceeds further and the elastic block is moved away
from the contact, the shape of the deformed profile significantly changes
compared to the loading case (see $s=51,92\mathrm{\mu m}$), and is
characterized by a significantly larger contact area and by the formation
stretched contacts with pronounced adhesive necks \cite{Zappone-2007}. This
type of behavior is peculiar of asperity adhesive contact \cite{Johnson-1971}.
In fact, in presence of adhesion, asperities enter in contact when the local
interfacial load is still zero, but during unloading asperities are first
stretched, with the formation of adhesive neck, and then jump out of contact
at negative local loads. During unloading unstable local pull-off events occur
at random locations \cite{Zappone-2007} leading to pronounced differences
between the loading and unloading precesses. Such different behaviors can be
indirectly observed in in Fig \ref{04}, where the PSD of the rigid substrate
profile (fractal dimension $D_{\mathrm{f}}=1.2$, non dimensional \ penetration
$\tilde{\Delta}=\Delta/h_{\max}=0.4$) has been compared to the PSD\ of the
deformed shape of the elastic body during loading and unloading conditions,
for $\gamma=0.01\mathrm{J/m}^{2}$ [Fig. \ref{04} (b)] and $\gamma
=0.04\mathrm{J/m}^{2}$\ [Fig. \ref{04} (b)].

\begin{figure}[h]
\begin{center}
\includegraphics[width=8.00cm]{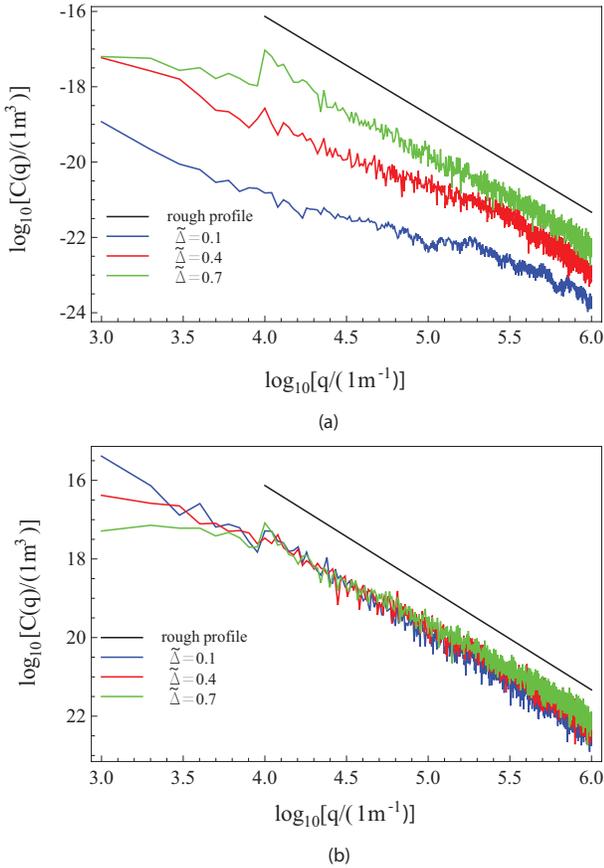}
\end{center}
\caption{(Color on line) The PSD of the rigid substrate profile with fractal
dimension $D_{\mathrm{f}}=1.2$, compared to the PSD\ of the deformed shape of
the elastic body, obtained for $\gamma=0.04\mathrm{J/m}^{2}$, (a) for loading
and (b) unloading conditions, for different values of the non dimensional
penetration $\tilde{\Delta}=\Delta/h_{\max}$.}%
\label{05}%
\end{figure}

\begin{figure}[ptb]
\begin{center}
\includegraphics[width=8cm]{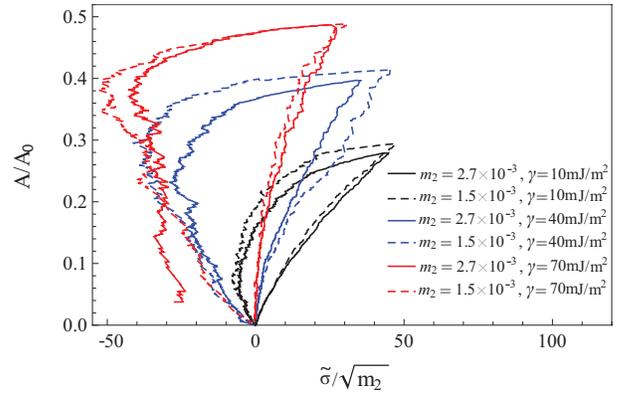}
\end{center}
\caption{(Color on line) The true contact area $A/A_{0}$ as a function of the
dimensionless quantity $m_{2}^{-1/2}\tilde{\sigma}$ for two different Hurst
exponents $H=0.8$ ($m_{2}=2.7\times10^{-3}$), $H=0.9$ ($m_{2}=1.5\times
10^{-3}$), and for three different values of energy of adhesion,
$\gamma=0.01\mathrm{J/m}^{2}$ (black curves), $\gamma=0.04\mathrm{J/m}^{2}$
(blue curves), and $\gamma=0.07\mathrm{J/m}^{2}$ (red curves). As predicted by
the theories, there is a marginal influence of the fractal dimension on the
true contact area at small loads under loading conditions. The influence of
$D_{f}$ becomes important during unloading. This leads to the formation of a
hysteresis loading- unloading loop, which is strongly affected by the adhesion
energy $\gamma.$}%
\label{06}%
\end{figure}

At first we remark that for large wavevectors $q$, the PSDs of the
deformed profile, either in loading or unloading conditions\ (blue and red
curves respectively), run parallel to the PSD of the rigid rough profile. This
is due to the fact that, for $0.5<H<1$, full contact always occurs between the
elastic block and the short wavelength corrugation of the rough rigid profile,
(see Fig. \ref{03}). In fact, we can easily estimate the threshold wavelength
$l_{th}$ below which full contact occurs between solid and the rigid rough
substrate. To this end consider that for fractal surface the amplitude
$A\left(  q\right)  =2\left\vert a\left(  q\right)  \right\vert $ of each
single wavy corrugation scales as $A\left(  q\right)  /A\left(  q_{\min
}\right)  \sim\left(  q_{\min}/q\right)  ^{H}$ where $A\left(  q_{\min
}\right)  $ is of order of the rms roughness $h_{\mathrm{rms}}$ of the
substrate. Now assume that the elastic slab makes contact with the surface on
a region of size $l=2\pi/q$ in this case, assuming $A\left(  q\right)  \ll l$,
the change of elastic energy stored in the body can be shown to be $\Delta
U_{el}\sim El^{2}A\left(  q\right)  $ whereas the change of adhesion energy
upon contact is $\Delta U_{ad}\sim-\gamma l^{2}$. Therefore full contact will
occur when the change of total energy $\Delta U_{tot}=\Delta U_{el}+\Delta
U_{ad}$ upon contact is $\Delta U_{tot}<0$. In this case the contact will
occur on a single connected region, otherwise it will be split in many
different contact spots. The condition $\Delta U_{tot}<0$ gives $EA\left(
q\right)  <\gamma$, i.e. $A\left(  q\right)  <\delta=\gamma/E$ where $\delta$
is called the adhesion length. Using $A\left(  q\right)  /A\left(  q_{\min
}\right)  \sim\left(  q_{\min}/q\right)  ^{H}$ one gets $l<\left(
2\pi/q_{\min}\right)  \left[  \delta/A\left(  q_{\min}\right)  \right]
^{1/H}=l_{th}$. In our case we get $l_{th}\approx2-12\mathrm{\mu m}$ depending
on the value of the energy of adhesion. Therefore, during loading, at scales
below this threshold value $l_{th}$ we will observe the formation of small
contact where the elastic solid conforms to the rigid substrate, thus leading
to the observed trend of PSD\ (see Fig. \ref{04}), which, indeed, runs
parallel to the PSD\ of the rigid rough profile. At smaller spatial
frequencies the contact is, instead, split in many different disconnected
regions. Therefore partial contact occurs at the large scales and the slope of
the PSD of the deformed profiles must necessarily differ from the one of the
rigid rough profile. In \cite{Carbone-2012} the authors have shown that, under
load conditions, in such smaller range of spatial frequencies and during
loading conditions, the PSD\ of the deformed profile closely follows a power
law of the type $C_{v}(q)\approx q^{(2+H)}$ in very good agreement with
Persson's theory \cite{Persson-2002, Persson-2005, Yastrebov-2012, Dapp-2015,
Prodanov-2014, Dapp-2012, Dapp-2014}. However, during unloading the PSD of the
deformed profile shows a different trend: (i) for large wave-vectors the
unloading PSD still runs parallel to the PSD of rough profile but over a wider
range of wave-vectors, (ii) at smaller spatial frequencies the slope of the
PSD, instead, does not follow Persson's predictions, as shown by the larger
slope of the PSD curve compared to the loading case. This fact seems to
suggest that the power law trend predicted by Persson's theory
\cite{Persson-2002, Persson-2005} holds true only for loading conditions. This
is more evident at larger values of adhesion energy, see curve for
$\gamma=0.04\mathrm{J/m}^{2}$ in Fig \ref{04} (b). When the PSDs of the
deformed profile are compared at different values of the penetration (see Fig.
\ref{05}), an interesting different behavior between loading and unloading can
be observed. In particular, we note that, during loading, the PSD\ of the
deformed surface is very sensitive to the penetration value $\tilde{\Delta}$
[Fig.\ref{05} (a)], whilst this sensitivity is much less pronounced during
unloading [Fig.\ref{05} (b)], at least for relatively large values of the
energy of adhesion $\gamma=0.04\mathrm{J/m}^{2}$.In Fig. \ref{06} the normalized real contact area $A/A_{0}$ is
shown, for different values of average square slope of the profile
$m_{2}=\left\langle h^{\prime2}\right\rangle $ and adhesion energy $\gamma$,
as a function of the dimensionless quantity $\tilde{\sigma}/\sqrt{m_{2}}$,
where $\tilde{\sigma}=2\sigma/(E^{\ast}q_{0}h_{\max})$. We have chosen to plot
$A/A_{0}$ vs. $\tilde{\sigma}/\sqrt{m_{2}}$ as, for adhesiveless contacts,
theories and numerical calculations [7,10,17,18,34] predict an almost linear
relation between this two quantities, which is observed to be independent of
the elastic properties of the material and the surface
statistics.\begin{figure}[ptb]
\begin{center}
\includegraphics[width=8.00cm]{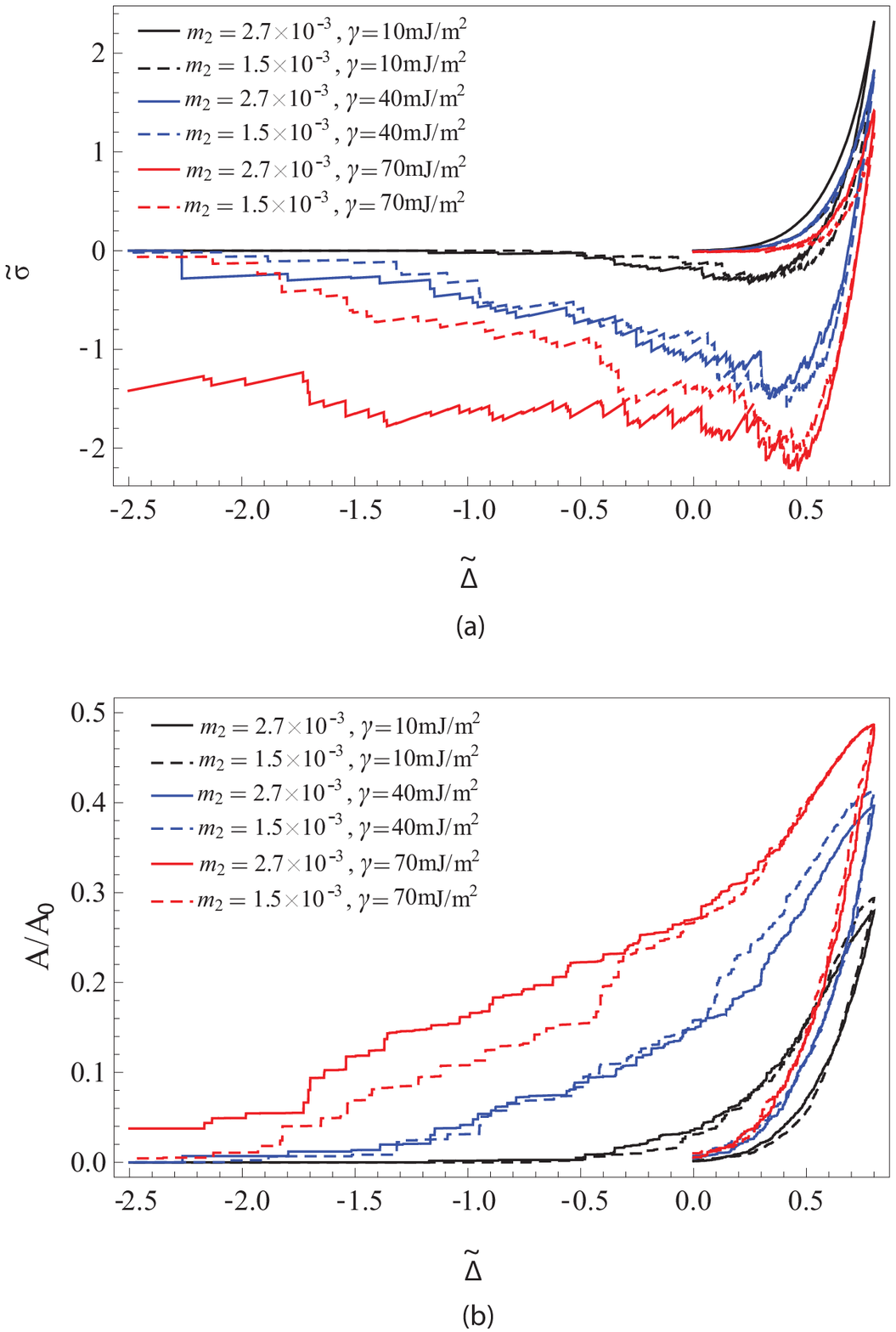}
\end{center}
\caption{(Color on line) The dimensionless load $\tilde{\sigma}=2\sigma
/(E^{\ast}q_{0}h_{\max})$ (a) and the quantity $A/A_{0}$ (b) as a function of
the non dimensional \ penetration $\tilde{\Delta}$, for two different Hurst
exponents $H=0.8$ ($m_{2}=2.7\times10^{-3}$), $H=0.9$ ($m_{2}=1.5\times
10^{-3}$), and for three different values of energy of adhesion,
$\gamma=0.01\mathrm{J/m}^{2}$ (black curves), $\gamma=0.04\mathrm{J/m}^{2}$
(blue curves), and $\gamma=0.07\mathrm{J/m}^{2}$ (red curves).}%
\label{07}%
\end{figure}\begin{figure}[ptb]
\begin{center}
\includegraphics[width=8.00cm]{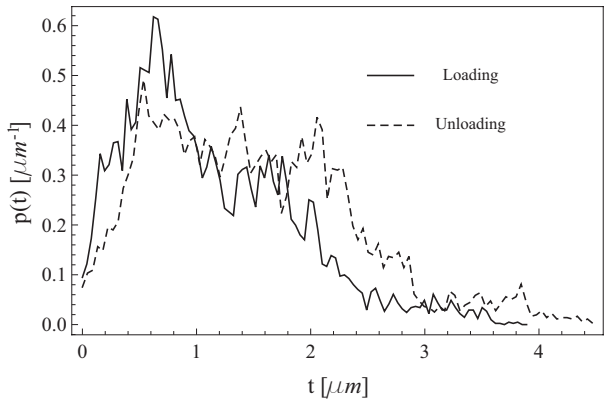}
\end{center}
\caption{The normalized probability density function (PDF) $p\left(  t\right)
$ of the local separation $t\left(  x\right)  $ in the non-contact area
($t\left(  x\right)  >0$) for $H=0.7$, and $\gamma=0.04\mathrm{J/m}^{2}$,
$\tilde{\Delta}=0.7$ for (a) loading and (b) unloading conditions.}%
\label{08}%
\end{figure}However, in our case, this linearity is not observed, especially
when the two surfaces are moved apart (unloading). More importantly, given the
same applied load, the contact area during unloading is much larger than
during loading, thus leading to strong hysteresis. This is even more clear in
Figure \ref{07} (a) where the dimensionless load $\tilde{\sigma}$ is plotted
vs. the non dimensional penetration $\tilde{\Delta}$. Interestingly, being the
solid perfectly elastic, such a hysteretic energy dissipation must be related
to contact phenomena occurring at the interface. In fact we can propose two
different mechanisms leading to energy dissipation: (i) the first occurring at
small scales, i.e. for wavelength $l<l_{th}$, which we refer to as the
Small-Scale-Hysteresis (SSH); (ii) the second at large scales, i.e. for
$l>l_{th}$, which we refer to as the Large-Scale-Hysteresis (LSH). To
understand the origin of the SSH, let us recall that at small scales the solid
conforms to the rigid substrate. Thus, each single contact is actually
represented by a compact interval. In such a case Guduru and his collaborators
have shown that, already a moderate roughness strongly modifies the original
JKR curve providing it with a non-monotonic behavior \cite{Guduru-2007,
Guduru-2007BIS}. This, in turn, causes the unloading process to be
characterized by many crack propagation jumps, with the interface separating
in alternating stable and unstable segments. Each unstable segments dissipates
energy leading to an increase of the total work during unloading and, hence,
to energy dissipation. This unstable behavior has been used in Ref.
\cite{Kesari-2010} to justify the hysteresis observed in JKR experiments with
AFM tips.

The origin os LSH is instead different. In fact, as noted so far, when the
surface is observed at large scales the contact is constituted by a set of
disconnected small contact regions, where the short wavelength corrugation is
apparently smoothed out. The contact interface, then, appears as the contact
between as certain number of randomly located smooth asperities, each of one
obeying the JKR adhesion laws. In such conditions Israelachivili and his group
\cite{Zappone-2007} proposed a very simple picture to explain the occurrence
of hysteresis. As noted so far, this is indeed due to the local stretching and
consequent JKR\ pull-off of asperities during unloading \cite{Zappone-2007}.

It is noteworthy to observe that increasing the energy of adhesion $\gamma$
leads to a strong increase of hysteresis loop [see Fig. \ref{07} (a)]. The
origin of this behavior is twofold as increasing $\gamma$ necessarily leads
to: (i) an increase of number of contact patches, and (ii) to an increase of
the size of each single contact patch. This, in turn, determines an
enhancement of SSH and LSH phenomena, i.e. to a large increment of energy
dissipation during the loading-unloading loop. Figure \ref{07} (b) shows the
reduced real contact area $A/A_{0}$ as a function of the dimensionless
penetration $\tilde{\Delta}$ for two different Hurst exponents $H=0.8$
($m_{2}=2.7\times10^{-3}$), $H=0.9$ ($m_{2}=1.5\times10^{-3}$), and for three
different values of energy of adhesion, $\gamma=0.01\mathrm{J/m}^{2}$ (black
curves), $\gamma=0.04\mathrm{J/m}^{2}$ (blue curves), and $\gamma
=0.07\mathrm{J/m}^{2}$ (red curves).\bigskip%

\begin{tabular}
[c]{lcll}
& $\gamma=0.01\mathrm{J/m}^{2}$ & $0.04\mathrm{J/m}^{2}$ & $0.07\mathrm{J/m}%
^{2}$\\
$m_{2}=5.1\times10^{-3}$ & - & 26.6 & 42.83\\
$2.7\times10^{-3}$ & 3.44 & 19.36 & 49.25\\
$1.5\times10^{-3}$ & 2.59 & 17.91 & 28.15
\end{tabular}

{\small Tab.1 - The dimensionless energy loss. }\bigskip

An estimation of the dimensionless energy loss during the entire
loading-unloading cycle, for different values of the average square slope of
the profile, is given in Table 1. In particular no general trend can be
inferred. Results, instead, may suggest a non monotonic dependence of the
dissipated energy on $m_{2}$. This is, infact, what we expect. Simple
dimensional argument show indeed that the density $D_{\mathrm{sum}}$ of the
substrate asperities increases as $D_{\mathrm{sum}}\sim\sqrt{m_{4}/m_{2}}$
where $m_{4}=\left\langle h^{\prime\prime2}\right\rangle =\int q^{4}%
C_{R}\left(  q\right)  dq$ is the average square curvature of the rough
profile. Thus, considering that $m_{4}$ increases with $m_{2}$ faster than
$m_{2}$, it follows that the number of asperities per unit length grows as
$m_{2}$ is increased. Therefore, for moderate values of the average square
slope, increasing $m_{2}$ should lead, by following the mechanism proposed in
Ref. \cite{Guduru-2007, Guduru-2007BIS} to an enhancement of the SSH
hysteresis in each contact spot, and to and increment of the number of contact
spots, thus increasing the number of LSH pull-off events. Hence, for small
$m_{2}$, increasing $m_{2}$ should necessarily causes an increase of hysteresis.

However, for sufficiently large values of $m_{2}$, the number of contact spots
as well as the contact area in each contact spot must strongly diminish (given
the same value of $\gamma$). This follows from the fact that, as shown above,
full contact between the elastic solid and the rigid substrate on the
length-scale $l=2\pi/q$, occurs when the amplitude $A\left(  q\right)  $ of
the spectral components satisfies the relation $A\left(  q\right)  <\delta$,
that is to say $m_{2}\left(  q\right)  <\left(  \delta/l\right)  ^{2}$, where
$m_{2}\left(  q\right)  \approx\left[  A\left(  q\right)  /l\right]  ^{2}$ is
the average square slope of $q$-component of the rigid rough profile.
Therefore, increasing $m_{2}$ will cause full contact conditions to be
established at continuously decreasing lengthscales. This will strongly reduce
the contact area in each contact spots, will reduce the force needed to detach
the elastic body from the rigid substrate and in the end strongly diminish the
adhesion hysteresis, in agreement with experimental observations in AFM
contacts in Ref. \cite{Kesari-2010}.

Fig. \ref{08} shows the normalized conditional probability density function
(PDF) $p\left(  t\right)  $ of the interfacial separation $t\left(  x\right)
=v\left(  x\right)  -h\left(  x\right)  +s$ in the non contact area (where
$t>0$). $t\left(  x\right)  $ plays a crucial rule in many practical
applications (e.g. mixed lubrication, lip seals, static seals). The PDF\ has
been calculated for $H=0.7$, and $\gamma=0.04\mathrm{J/m}^{2}$, in (a) loading
and (b) unloading conditions. The trend of the calculated $p\left(  t\right)
$ in the loading and unloading conditions slightly differs in Figure \ref{08},
since during unloading (see Fig. \ref{03}), the shape of the deformed profiles
changes leading to higher interfacial separations $t\left(  x\right)  $
compared to the loading case. This explains the trend of $p\left(  t\right)  $
in Figure \ref{08} for the unloading case: the $p\left(  t\right)  $ is, in
fact, slightly shifted towards higher values of separation $t\left(  x\right)
$ compared to the loading case.

\section{Conclusions}

In this paper we have studied the adhesive contact between a rubber block and
a rigid randomly rough profile during loading and unloading conditions. The
roughness has been considered to be a self-affine fractal on a limited range
of wave-vectors. Calculations have been carried out for each profile by means
of an ad hoc numerical code previously developed by the authors. The
calculated data have been statistically averaged, and the influence of profile
average slope and energy of adhesion on loading-unloading contact behavior has
been investigated. We have shown that the combination of adhesion and
roughness leads to the appearance of hysteresis cycle and, hence, to energy
dissipation. We physically justify the observed behavior by considering two
sources of energy dissipation one occurring at small scales and the second at
large scales. We have numerically quantified the energy loss depending on the
average slope of roughness and on the energy of adhesion and discuss it in
terms of surface statistical properties.

\begin{figure}[ptb]
\begin{center}
\includegraphics[width=8cm]{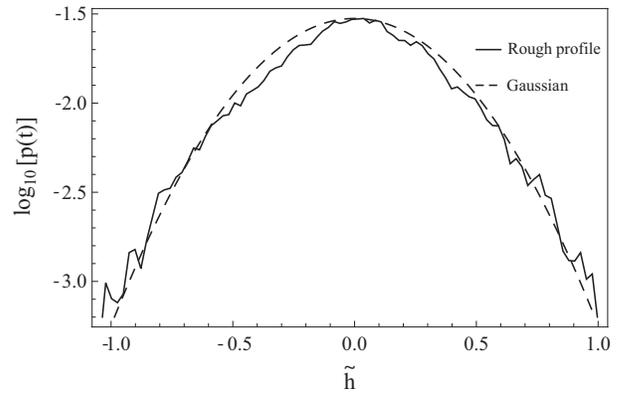}
\end{center}
\caption{The averaged dimensionless height probability density function
$p\left(  t\right)  $ of 10 rigid rough profiles. $\tilde{h}\left(  x\right)
=h\left(  x\right)  /h_{\max}$. Calculations are shown for $H=0.8$. Because of
the enhanced ergodicity, the trend of the calculated probability density
function $p\left(  t\right)  $ (solid line) follows pretty well a Gaussian
random distribution (dashed line).}%
\label{09}%
\end{figure}

\appendix

\section{Rough profile generation \label{appendice}}

In our numerical calculations we have utilized a periodic profile with
wavectors in the range $q_{0}<q<q_{\max}$. In particular, the non-vanishing
spectral components of our profiles are given by Eq. \ref{PSD} in the range
from $q_{\min}=n_{0}q_{0}<q<q_{\max}=Nq_{\min}$. Out of this range the PSD is
zero. This choice is necessary in order to improve the ergodicity of the
process. For the numerical generation of a profile, it is necessary to
determine the amplitudes $\left\vert a_{m}\right\vert $ and the phases
$\phi_{m}$ of the terms $a_{m}=\left\vert a_{m}\right\vert e^{i\phi_{m}}$ [see
Eq. (\ref{general rough profile})]. It can be shown that in order to satisfy
the translational invariance of the profile statistical properties (which
implies that the autocorrelation function satisfies the relation $\left\langle
h\left(  x^{\prime}\right)  h\left(  x^{\prime}+x\right)  \right\rangle
=\left\langle h\left(  0\right)  h\left(  x\right)  \right\rangle $), it is
enough to assume that the random phases $\phi_{m}$ are uniformly distributed
on the interval $\left[  -\pi,\pi\right[  $. This also guarantees that the
process is Gaussian. Now moving from Eq. (\ref{general rough profile}) the
PSD\ is
\begin{equation}
C\left(  q\right)  =%
{\displaystyle\sum\limits_{m=-\infty}^{+\infty}}
\left\langle \left\vert a_{m}\right\vert ^{2}\right\rangle \delta\left(
q-mq_{0}\right)
\end{equation}
from which it follows%
\begin{equation}
C\left(  mq_{0}\right)  =\left\langle \left\vert a_{m}\right\vert
^{2}\right\rangle \delta\left(  0\right)
\end{equation}
If we assume self-affine fractal profile [see Eq. (\ref{PSD})] one obtains
\begin{equation}
\left\langle \left\vert a_{m}\right\vert ^{2}\right\rangle =\left\langle
\left\vert a_{1}\right\vert ^{2}\right\rangle m^{-\left(  2H+1\right)  }%
\end{equation}
Hence, the quantity $\left\langle \left\vert a_{m}\right\vert ^{2}%
\right\rangle $ can be determined once known $\left\langle \left\vert
a_{1}\right\vert ^{2}\right\rangle $ and the Hurst exponent of the
surface.\ However to completely characterize the rough profile we still need
the probability distribution of the amplitudes $\left\vert a_{m}\right\vert $.
There are several choices, however the simplest assumption, as suggested by
Persson et al. in Ref. \cite{Persson-2005}, is that the probability density
function of $\left\vert a_{m}\right\vert $ is just a Dirac's delta function
centered at $\left\langle \left\vert a_{m}\right\vert ^{2}\right\rangle
^{1/2}$, i.e.
\begin{equation}
p\left(  \left\vert a_{m}\right\vert \right)  =\delta\left(  \left\vert
a_{m}\right\vert -\left\langle \left\vert a_{m}\right\vert ^{2}\right\rangle
^{1/2}\right)
\end{equation}
Fig. \ref{09} shows the averaged probability density function $p\left(
t\right)  $ of 10 profile dimensionless roughness heights $\tilde{h}\left(
x\right)  =h\left(  x\right)  /h_{\max}$. The Hurst exponent is $H=0.8$.
Thanks to the improved ergodicity of the numerically generated rough profile,
the trend of the calculated probability density function $p\left(  t\right)  $
(solid line) follows closely a Gaussian distribution (dashed line).

\end{document}